\documentclass[10pt]{article}

\usepackage{amsmath}
\usepackage{amssymb}

\usepackage{cite}

\usepackage{color} 

\usepackage[labelfont=bf,labelsep=period,justification=raggedright]{caption}

\makeatletter
\renewcommand{\@biblabel}[1]{\quad#1.}
\makeatother

\date{}

 \usepackage[pdftex]{graphicx}
\DeclareGraphicsExtensions{.pdf,.jpeg,.jpg,.png}

\begin{document}

\begin{flushleft}
{\Large
\textbf{A validated model of serum anti-M\"{u}llerian hormone from conception to menopause.}
}
\\
Thomas W Kelsey$^{1}$, 
Phoebe Wright$^{2}$, 
Scott M Nelson$^{3}$, 
Richard A Anderson$^{4}$, 
W Hamish B Wallace$^{4\ast}$
\\
\bf{1}  School of Computer Science, University of St. Andrews, St. Andrews, Scotland, UK
\\
\bf{2} University of Edinburgh,  Edinburgh, Scotland, UK
\\
\bf{3} Centre for Population and Health Sciences, University of Glasgow, Glasgow, Scotland, UK
\\
\bf{4}  Division of Reproductive and Developmental Sciences, University of Edinburgh,  Edinburgh, Scotland, UK
\\
$\ast$ E-mail: Corresponding hamish.wallace@nhs.net
\end{flushleft}

\section*{Abstract}
{\it Background}: 
Anti-M\"{u}llerian hormone (AMH) is a product of growing ovarian follicles. The concentration of AMH in blood may also reflect the non-growing follicle (NGF) population, i.e. the ovarian reserve, and be of value in predicting reproductive lifespan. A full description of AMH production up to the menopause has not been reported.

\noindent{\it Methodology/Principal Findings}:
By searching the published literature for AMH concentrations in healthy pre-menopausal females, and using our own data  (combined $n=3,260$) we have generated and robustly  validated the first model of AMH concentration from conception to menopause. This model  shows that 34\% of the variation in AMH  is due to age alone. We have shown that AMH peaks at age 24.5 years, followed by a decline to the menopause. We have also shown that there is a neonatal peak and a potential pre-pubertal peak. 
Our model allows us to generate normative data at all ages.

\noindent{\it Conclusions/Significance}:
These data highlight key inflection points in ovarian follicle dynamics.  This first validated model of circulating AMH in healthy females describes a transition period in early adulthood, after which AMH reflects the progressive loss of the NGF pool. The existence of a neonatal increase in gonadal activity is confirmed for females. An improved understanding of the relationship between circulating AMH and age will lead more accurate assessment of ovarian reserve for the individual woman.

\section*{Author Summary}
Women are born with their full compliment of immature eggs, that declines with increasing age to the menopause at an average age of 50 years. There is currently no accepted test that will reliably predict how many immature eggs remain for an individual woman (their ovarian reserve). This is of particular importance to young women who may be at risk of a premature menopause after cancer treatment or who may be considering assisted reproduction. Serum anti-M\"{u}llerian hormone (AMH) is produced by cells that surround the developing and maturing eggs. The measurement of serum AMH is currently used by many clinicians as a surrogate measure of ovarian reserve.  However, little is known about how reliably AMH reflects ovarian reserve and activity for the individual woman. In this study, using out own data and data previously published from many sources in healthy women, we have provided the first rigorously validated model of AMH from conception to menopause. We have shown for the first time that AMH peaks at age 24.5 years, followed by a decline to the menopause when it is undetectable. This model allows us to generate normal range values for AMH from birth to menopause. An improved understanding of the relationship between circulating AMH and age will allow clinicians to more reliably assess ovarian reserve, and ultimately allow women to plan their own reproductive course with confidence.

\section*{Introduction}
The human ovary establishes its complete complement of primordial follicles during fetal life, with recruitment and thereby depletion of this dormant primordial follicle pool required for normal fertility but ultimately leading to reproductive senescence \cite{Broekmans09}. Primordial follicles are recruited continuously from before birth to join the early growing cohort (initial recruitment). After puberty, at every new cycle a limited number of follicles are recruited from this cohort of small growing follicles (cyclic recruitment), followed by a final selection for dominance and ovulation of a single follicle  \cite{Gougeon96,McGee00}. Thus, at any specific time, the majority of primordial follicles are held in a dormant state, and when eventually recruited most will not reach the preovulatory stage but are destined to be removed through atresia at earlier stages of follicular development. Currently, clinical assessment is unable to assess the number of primordial follicles, or their rate of loss/activation. Knowledge of these aspects of ovarian function would be of value in a range of contexts, both clinical and social/personal, as well as being of great value in promoting our understanding of how reproductive lifespan is regulated.

Anti-M\"{u}llerian hormone (AMH), is now recognized as a principal regulator of early follicular recruitment from the primordial pool \cite{Durlinger02,Gigli05}, with AMH null mice demonstrating accelerated depletion of primordial follicle number and an almost three-fold increase in smaller growing follicles \cite{Durlinger99}.  Furthermore this increase in number of growing follicles occurs despite  lower serum follicle stimulating hormone (FSH) concentrations \cite{Durlinger01}, suggesting that in the absence of AMH, follicles are more sensitive to FSH and progress through the early stages of follicular development.  AMH is produced by small growing but not  primordial follicles \cite{Weenen04,Bezard87,Baarends95}, although limited data suggest that serum AMH concentrations also correlate with primordial follicle number in humans \cite{Hansen10} as in rodents \cite{Kevenaar06}.   The prepubertal endocrine environment is markedly different from the adult with low and non-cyclical gonadotropins: the relevance of this to AMH secretion is incompletely understood although follicle growth through the preantral stages and occasionally to early antral stages (i.e. across the full range of stages that secrete AMH)  is observed in childhood \cite{Lintern-Moore74}.  
A recent study has reported an increase in initial primordial follicular recruitment rates up to the age of puberty, and then a progressive decline to the menopause\cite{WK}.  This suggests that AMH concentrations at any given age in both childhood and adulthood may mirror primordial follicular recruitment rates, rather than simply primordial follicle number. Consequently across the female lifespan, circulating AMH will potentially exhibit an initial increase followed by a non-linear decline as is well established for the primordial follicle pool \cite{Faddy92,Faddy96,Faddy00,Hansen08,WK}.

In keeping with this, AMH concentrations in adults have been shown to decline with age \cite{DeVet2002, VanRooij20 05}.  AMH concentrations are relatively stable across the menstrual cycle \cite{Cook00,LaMarca04} and also between cycles in the same woman \cite{Fanchin05}.  Measurement of AMH is increasingly used in the prediction of ovarian response to superovulation  \cite{LaMarca10}. 
Although  large AMH cohort studies describing falling AMH concentrations in adult women (mostly from populations attending infertility clinics) have recently been published \cite{Nelson10,Nelson11,Seifer11,Almog11}, the data for AMH concentrations in children have until recently been considerably more limited \cite{Hagen2010,Ahmed2010}. To date no single study has examined AMH across the lifespan in healthy females.  The aim of the current study is  to produce a model of serum AMH in healthy females from conception to the menopause.

\section*{Results}

\subsection*{The validated model}

\begin{figure}[!ht]
\begin{center}
\includegraphics[width=6.5in]{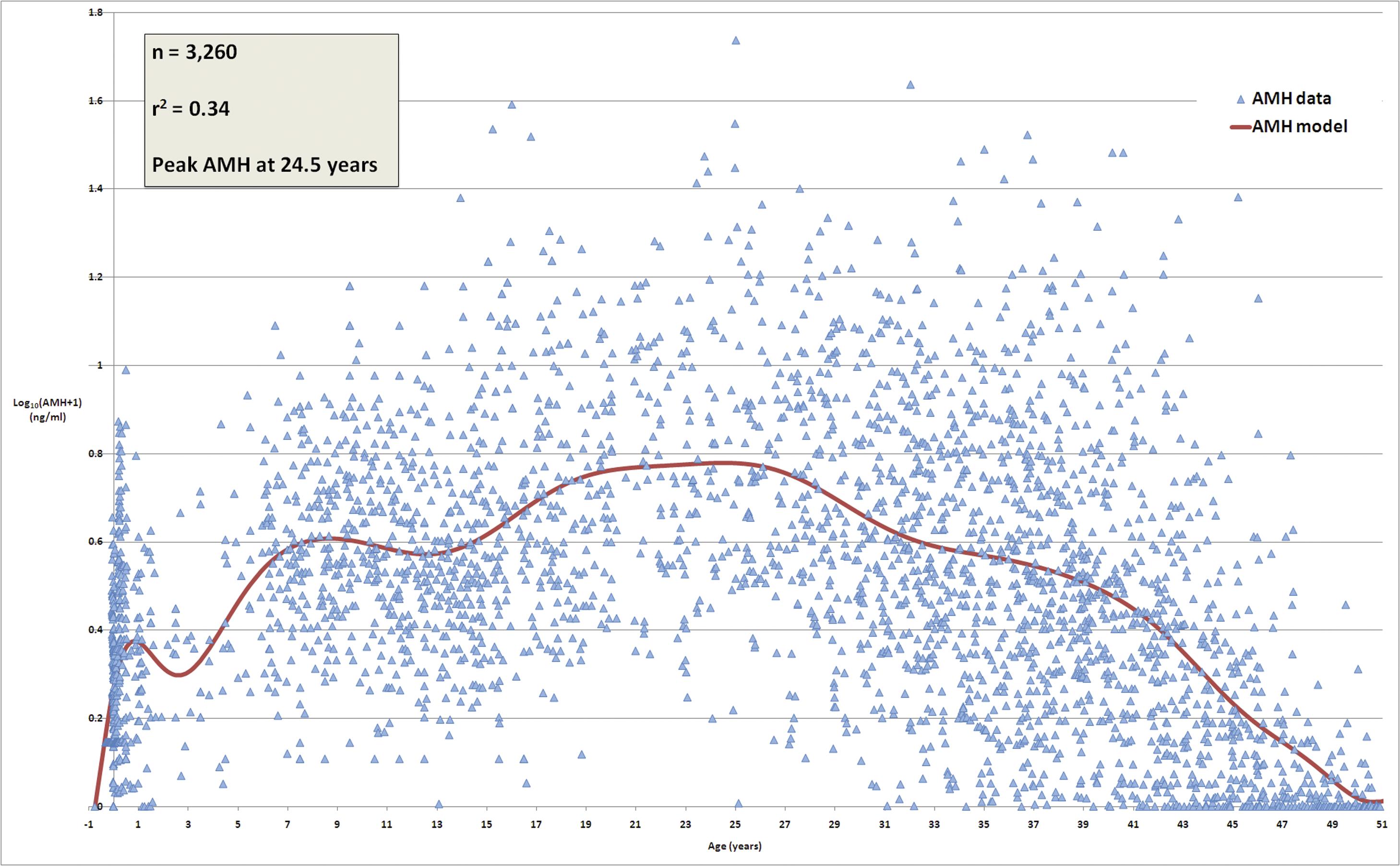}
\end{center}
\caption{
{\bf  Serum AMH data}  The red line is the model that best fits the 3,260 datapoints shown as triangles. The coefficient of determination, $r^2$, is 0.34, indicating that 34\% of variation in serum AMH concentrations is due to age alone. Peak serum AMH is at 24.5 years. }
\label{Figure_01}
\end{figure}

Data  from  published studies and our own  (Table 1) were used to derive the model (Figure 1).  This included 3,260 data points across the age range from -0.3 (cord blood from preterm infants) to 54.3 years (Table 2). After 10-fold stratified cross-validation, the model with the highest coefficient of determination ($r^2$) was a degree 20 polynomial of the form

$$
log_{10}(AMH + 1) = c_0 + c_1\mathrm{age}  + c_2\mathrm{age}^2 + \ldots +  c_{20}\mathrm{age}^{20}
$$
with coefficients $c_j$ given in Table 3.
 
 This model has an $r^2$ of 0.34, indicating that roughly one third of variation in AMH concentrations is due to age alone, with the remaining two thirds of the variation being due to other factors.   Serum AMH peaks at 24.5 years on average,  with concentrations decreasing shortly after birth, and again between eight and twelve years of age. 

 \subsection*{Validation}

For each of the 10 folds of the validation process, the highest ranked of 215 models was a polynomial similar to the model reported above: in every case the returned coefficients, $r^2$ and peaks were  similar (Table 4). Moreover, the average training error was within 1\% of  the average cross-validation estimate of the prediction error across 10 folds (Table 4).  We  therefore consider that the model derived for all 3,260 datapoints generalises well to unseen data, and hence report this as a validated model for serum AMH concentrations in the normal female population (Figure 2).


%

\section*{Discussion}

 We have generated the first validated model of serum AMH in the healthy human population from conception to menopausal ages. Our model shows that serum AMH concentrations peak at age 24.5 years for the average case, and suggests that two thirds of the variation in AMH concentrations for healthy females is due to factors other than age.

 We have shown that serum AMH falls shortly after birth, with concentrations only increasing again after about two years of age. This feature is in line with evidence of a mini-puberty seen in neonatal girls \cite{Lee03,Chellakooty03}, although more clearly characterized in boys \cite{Andersson98}, and with a recent longitudinal study of AMH in female neonatal blood \cite{Hagen2010} (the data from which were included in our study). Our model also shows that serum AMH concentration falls between the ages of eight and twelve, before rising to a peak in the mid-twenties. This fall may be an artefact of our model derivation process rather than a true  reproductive biological event.  As the fall coincides with the initial increases in gonadotropin concentrations of early puberty, it is possible that it reflects changes in the proportions of follicles at different stages of growth with increasing numbers progressing to antral stages rather than becoming atretic early on \cite{McGee00}. AMH is produced by early growing follicles at all stages up to the early antral stage \cite{Weenen04} but it is unknown which follicle class contributes most to circulating concentrations.  The rising granulosa cell mass (and thus AMH production per follicle) will be balanced by progressively declining numbers of follicles at each stage of growth \cite{Gougeon96,McGee00}.
 
Our incomplete understanding of how early follicle recruitment and growth are regulated means that any such interpretation is speculative but our results both suggest this avenue of future research, and give useful indications of effect sizes and ages for the design of such investigations.

The increase in AMH during the postnatal period, supported by a recent longitudinal analysis in the first 3 months of life \cite{Hagen2010}, is likely to be analogous to the well established transient rise in testosterone and inhibin B in boys at that time\cite{Andersson98}.  This is likely to reflect the relatively high gonadotrophin concentrations that are present which will support more advanced follicle development than occurs in the remainder of the prepubertal period. Consistent with this, the ovary shows follicle growth to the early antral stage from birth  \cite{Lintern-Moore74}.
 The continuing rise in AMH through childhood is striking, and parallels rising follicle growth initiation from the very large pool at these ages \cite{WK}.  The rising AMH production would therefore act to limit follicle growth activation \cite{Durlinger99,Gigli05}, thus a point of inflection when follicle recruitment starts to slow, and which is followed by a decline in AMH concentrations is predictable: our data demonstrate the age at which this occurs. 
  
Our observed increase in AMH concentrations beyond the age of 12 years, is in contrast to the recent analysis by Hagen  {\it et al}. which suggested in a cross-sectional study that AMH did not change from age 8 to age 25 or indeed relative to pubertal stage \cite{Hagen2010}. Censoring our own data between these two ages would also suggest that AMH does not vary markedly, however, this would lose the power of the broader picture afforded by the entire age range analysis performed here. The lack of an increase beyond age 12 as suggested by Hagen {\it et al}., is in marked contrast to  current and previous studies, which have all suggested a peak in early adulthood.  A recent analysis of 9,601 infertility patients \cite{Nelson10} and a smaller study of 82 healthy subjects  both report that AMH concentrations decrease after age 20 \cite{DeVet2002}. Although two studies have reported peak AMH at age 31-33 \cite{VanDisseldorp2008,Soto2009}, these were substantially smaller incorporating 144 and 58 subjects and the minimum age was 14 years and 25 years respectively. This smaller sample size and lack of data from younger (especially neonatal) subjects may have skewed the reported peaks towards a higher age. 

We recognize that although our study has a number of limitations -- including the accuracy of the health status of subjects in the studies used and the dependence on accurate graphical presentation of published results -- it also has considerable strengths.  The dataset was derived from over 3,500 subjects with ages ranging from  minus  0.3 years to 68 years, the model was then  validated using standard mathematical techniques, with very good generalisation to unseen data for each of the 10 cross-validation steps.  We did not make any presumptions regarding the optimal fitting model, yet the optimal model exhibited a non-linear decline in AMH in adult life consistent with previous cross-sectional and longitudinal studies \cite{Nelson10,LaMarca2005,VanDisseldorp2008,Sowers09}, and the minipuberty seen in neonatal girls\cite{Lee03,Chellakooty03,Hagen2010}.     
This validated model can therefore be used to accurately interpret concentrations of serum AMH in females across the lifespan. The model is based entirely on cross-sectional data (the small amount of longitudinal data was treated as if cross-sectional). Further longitudinal validation in similarly large populations will provide additional confirmation. 

It should be noted that our methodological choices have no effect on the qualitative nature of our results. If we convert to DSL values instead of IBC, do not add one to the log-adjusted values for ease of exposition, do not censor at age 54.3, allow models with more parameters (or restrict to models with slightly fewer parameters),  use $k\times2$ or bootstrap cross-validation, we still obtain a validated model with similar peaks, $r^2$ and level of generalisation to unseen data.   
The current introduction of the GenII AMH assay (Beckman Coulter) uses the same standards as the IBC assay, thus we anticipate that the normative model presented (Figure 2) will be valid for values obtained using that system.

\begin{figure}[!ht]
\begin{center}
\includegraphics[width=6.5in]{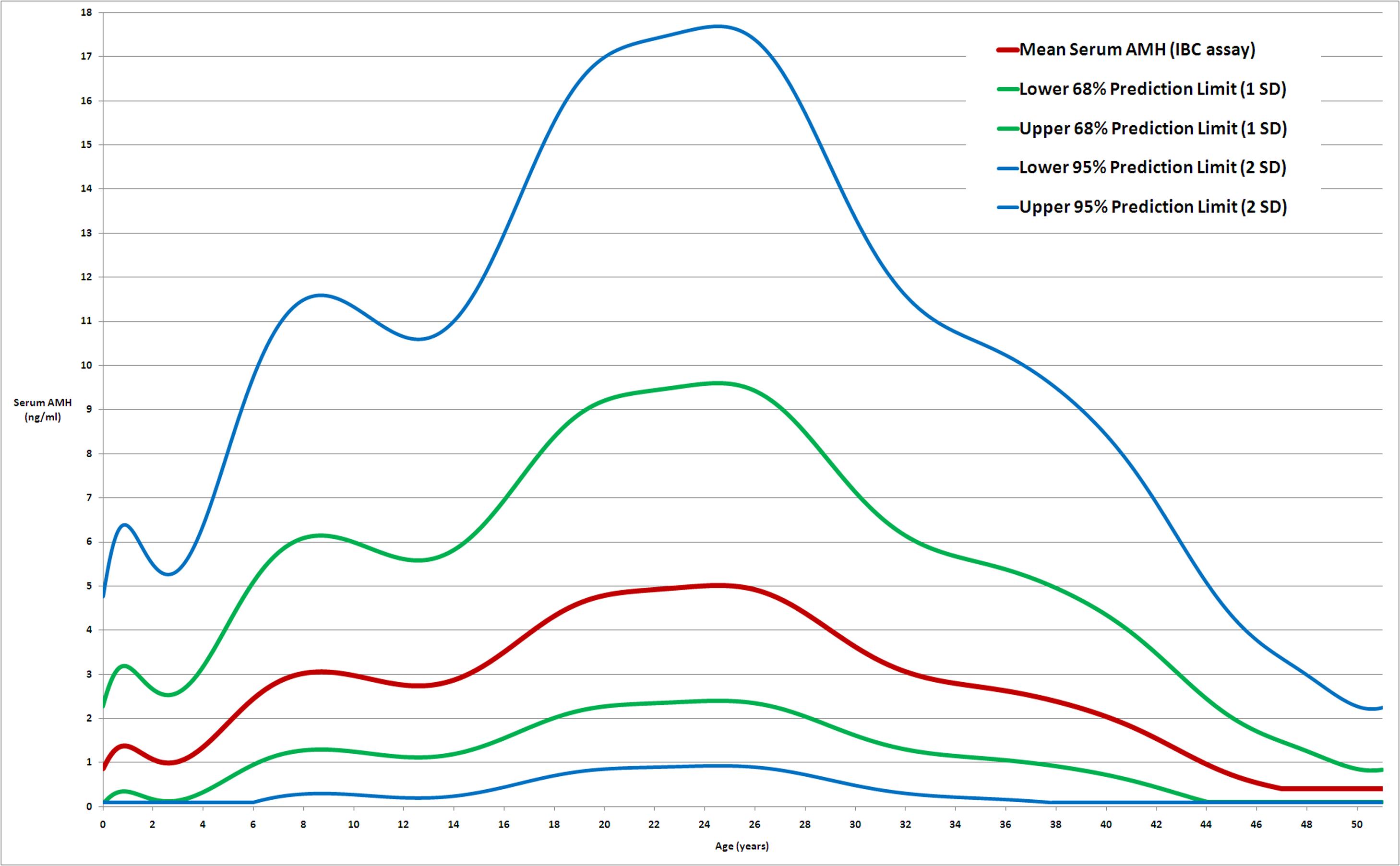}
\end{center}
\caption{
{\bf  The normal range for serum AMH in girls and women}  The red line is the log-unadjusted validated AMH model using IBC assay values. The blue and green lines are  the 68\% and 95\% prediction limits for the model (plus and minus one and two standard deviations respectively). }
\label{Figure_02}
\end{figure}

This comprehensive statistical analysis of  3,260 healthy infants, children and women has facilitated the first validated normative model of  age related circulating AMH from conception to the menopause. The model  provides a means for interpretation of how an individual's serum AMH concentration compares with population norms.

\section*{Materials and Methods}

\subsection*{Data acquisition}

\begin{table}[!ht]
\caption{
\bf{Serum AMH data summary}}
{\footnotesize
\begin{tabular}{|c|l|c|c|r|c|c|c|c|c|}
\hline
Ref. & $1^{st}$ Author & Data & Assay & $n$ & Average age & Age range & Det. lim.  & Intra CV & Inter CV\\ 
\hline
\cite{Soto2009}  & Soto & Graph & IBC & 58 & 30.3 (mean) & $\pm 8.7$ SD & 0.10 & 5.3 & 8.7 \\
\cite{Guibourdenche2003}  & Guibourdenche & Graph & IBC & 192 & NS & -0.3 -- 1.0 & 0.30 & 5.3 & 8.7 \\
\cite{Hudecova2009}  & Hudecova & Graph & IBC & 64 & 46.3 (mean)  & $\pm 6.4$ SD & 0.70 & 12.3 & 12.3 \\
\cite{Mulders2004}  & Mulders & Graph & IBC & 82 & 29.9 &  19.6 -- 35.6  & NS  & 5.0 & 8.0 \\
\cite{Pastor2005}  & Pastor & Graph & IBC & 42 &  NS & 18.0 -- 50.0  & 0.10 & 5.3 & 7.8 \\
\cite{Piltonen2005}  & Piltonen & Graph & IBC & 44 & 31.6 (mean) &  21.0 -- 44.0  & NS & 5.1 & 6.6 \\
\cite{VanRooij2005}  & van Rooij & Graph & IBC & 162 & NS & 25.0 -- 46.0  & 0.05 &   5.0  &  8.0 \\
\cite{Laven2004}  & Laven & Graph & IBC & 41 & NS  &  20.0 -- 36.0  & 0.05 & 5.0 & 8.0 \\
\cite{DeVet2002}  & de Vet & Graph & IBC & 82 & 29.0 & $\pm 4.0$ SD & 0.05 & 5.0 & 8.0\\
\cite{Knauff2009}  & Knauf & Graph & IBC & 83 & 34.2 (mean) & $\pm 3.4$ SD & 0.03 & 11.0 & 11.0 \\
\cite{Lee1996}  & Lee & Graph & IBC & 225 & NS  & 0.0 -- 51.0 & 0.50 & 9.0 & 15.0 \\
\cite{LaMarca2005}  & La Marca & Graph & IBC & 24 & 44.0  (mean) & $\pm 2.8$ SD & 0.24 & 5.0 & 8.0 \\
\cite{Hagen2010}  & Hagen & Graph & IBC & 891 & NS & 0.0--68.0 & 0.03 & 7.8 & 11.6 \\
\cite{VanBeek2007}  & van Beek & Graph & DSL & 82 &  29.0  & 20.0 -- 35.0 & NS  & 5.0 & 15.0 \\
\cite{Sanders2009}  & Sanders & Graph & DSL & 43 & 24.1 (mean) & 0.1 -- 51.0 & 0.01  & NS  & 11.4 \\
\cite{VanDisseldorp2008}  & van Disseldorp & Graph &DSL& 144 & 37.9 (mean) & 25.0 -- 46.0 & 0.03 & 11.0 & 11.0 \\
\cite{Tehrani2010}  & Tehrani & Graph & DSL& 267 & 27.1  & 16.0 -- 44.0  & 0.01 & 5.2 & 9.1 \\
\cite{Dorgan2009}  & Dorgan & Graph & DSL & 204 & 44.7 (mean) & 33.3 -- 54.7 & 0.06 & 8.0 & 8.0 \\
\cite{Ahmed2010}  & Ahmed & Raw &DSL & 128 & 8.5 &  0.5 -- 16.5  & 0.50 & 8.0 & 8.0 \\
\cite{Nelson10}      &    Nelson  & Raw & DSL & 441 &  36.1  & 21.9 -- 47.8  & 0.03 & 3.4 & 8.6 \\
\hline
\hline
    &  Total IBC & & & 1,990 & 15.8 & -0.3 -- 68.0 & & &  \\
            &  Total DSL& & & 1,309 & 35.4 & 0.2 -- 54.7 & & &  \\
            \hline
    &  Total $n$ & & & 3,299 &  34.0 & -0.3 -- 68.0 & & &  \\
                  \hline
   &  \bf{Censored total $\mathbf{n}$} & & & \bf{3,260} & \bf{28.3} & \bf{-0.3 -- 54.3} & & &  \\
                  \hline
\end{tabular}
}
\begin{flushleft} The references relate to the bibliography section of this paper.  Age information is given as median and range, or as mean and standard devation (SD), depending on which form was reported in the referenced study. Detection limits are given in ng/ml. Intra- and inter-observer coefficients of variation (CV) are percentages.  NS denotes not stated. The CVs are indications of the likely accuracy of repeated measurements being performed either by the same observer (intra), or by another observer with similar training (inter). For longitudinal studies -- \cite{Mulders2004,DeVet2002,Tehrani2010,Hagen2010} -- we report the average age of participants at  first measurement. The censored total excludes any values greater than 54.3 years (i.e. one standard deviation above the average age at menopause). 
\end{flushleft}
\label{tab:label}
 \end{table}

Studies involving serum AMH measurements of human females were identified by performing PubMed and Medline searches and searching individual journals (including Menopause, Fertility and Sterility, Human Reproduction and the Journal of Clinical Endocrinology and Metabolism) using the search terms AMH, M\"{u}llerian inhibiting substance, ovarian reserve and polycystic ovarian syndrome. The references of included studies (Table One; \cite{Soto2009,Guibourdenche2003,Hudecova2009,Mulders2004,Pastor2005,Piltonen2005,VanRooij2005,Laven2004,DeVet2002,Knauff2009,Lee1996,LaMarca2005,VanBeek2007,Sanders2009,VanDisseldorp2008,Tehrani2010,Dorgan2009,Hagen2010}) were checked to identify further relevant studies to be processed. Data was selected for this analysis only for subjects who were not known to be infertile, or have an identified chronic illness. Hence all subjects were either in  control groups from controlled studies or  from prospective studies of the healthy population. Any data from subjects with a chronic disease or undergoing infertility assessment or investigation were excluded from the study. In the main, the data was from pre-menopausal women. In three studies the menopausal status of the women was not stated \cite{Dorgan2009,Lee1996,Sanders2009}.
Data from fetal blood ($n=25$) and cord blood ($n=53$) of infants \cite{Guibourdenche2003} were included. Longitudinal data -- from \cite{Mulders2004,DeVet2002,Tehrani2010,Hagen2010} -- were recorded as cross-sectional values.
The  data were extracted from graphs using Plot Digitizer software \cite{PlotDigitizer} to convert datapoints on the graphs into numerical data. Repeated datapoints were isolated by requiring that the acquired dataset matched the descriptive statistics provided in the supporting paper (Table One). 

We combined the resulting dataset with two sets of raw Scottish data. The first consisted of  individual serum AMH measurements (n=441, median age 36.1 years, max. age 47.8, min age 21.9)  from a cohort of  women whose partners were known to have severe male factor infertility requiring ICSI, and where no other female cause of infertility had been identified  \cite{Nelson10}.  Individual patient serum AMH measurements were undertaken between July 2006 and October 2009 in the biochemical laboratories of the University of Glasgow, the Glasgow Centre for Reproductive Medicine and the Glasgow Royal Infirmary. All three facilities were providing centralised AMH testing for infertility clinics within the United Kingdom. Ethical approval for studies involving these data was obtained from NHS Scotland. Subjects were informed that data may be analysed anonymously. Individual ethical approval has not been taken for studies involving this data, as it is routine anonymous clinical data which is covered by the general UK National Health Service ethics for analysis of routine biochemical data, provided it is anonymous,
The second dataset was supplied by Ahmed et al. \cite{Ahmed2010} and consists of 128 measurements taken from subjects aged  0.5 -- 16.5 years.

 \begin{table}[!ht]
\caption{
\bf{Sample sizes at each age for the AMH model}}
{\footnotesize
\begin{tabular}{|c|r|c|r|c|r|c|r|c|r|c|r|}
\hline
 Age (yrs) & $n$ & Age (yrs) & $n$ & Age (yrs) & $n$ & Age (yrs) & $n$ & Age (yrs) & $n$ & Age (yrs) & $n$ \\ 
\hline
$\le 0$ & 144 &   &  &   &   &   &   &   &  &   &   \\
0 & 277 & 10 & 61 & 20& 21 & 30& 77 &40 & 79 & 50& 25 \\
1 & 43 & 11 & 69 & 21& 37& 31& 77 &41 & 72 & 51& 18 \\
2 & 12 & 12 & 65 & 22& 27 & 32& 101 &42 & 80 & 52& 19 \\
3 & 12& 13 & 61 & 23& 35 & 33& 74 &43 & 66 & $> 53$& 12 \\
4 & 14 & 14 & 82 & 24& 32 & 34& 97 &44 & 58 & &  \\
5 & 14 & 15 & 61 & 25& 50 & 35& 84 &45 & 59 &  &  \\
6 & 51 & 16 & 37 & 26& 36 & 36& 116 &46 & 36 & &  \\
7& 60 & 17 & 64 & 27& 55 & 37& 98 &47 & 37 & &  \\
8 & 60 & 18 & 40 & 28& 67 & 38& 100 &48 & 37 &  &  \\
9 & 58 & 19 & 40 & 29& 45 & 39& 86 &49 & 23 & &  \\
   \hline
\end{tabular}
}
\begin{flushleft}  The 3,260 AMH datapoints described in Table 1, split into ages.   $n$ denotes the size of the subset of the data associated with an age in years.
\end{flushleft}
\label{tab:label}
 \end{table}

 Serum AMH values were standardised to give AMH measurements in ng/ml using the conversion formula 1 pmol/l = 7.143 ng/ml.
 
The resulting data were considered separately depending on the assay used to obtain serum AMH values. The first dataset  (n=1,990, median age 15.8 years, max. age 68.0, min age -0.3) came from those studies in which the serum concentrations of AMH were determined using enzyme-linked immunoassay kits IBC (Immunotech Beckman Coulter Company, France).  The second dataset  (n=1,309, median age 35.4 years, max. age 54.7, min age 0.2) came from studies in which the enzyme-immunometric assay Active MIS/AMH ELISA kits DSL (Diagnostic Systems Laboratories Inc., TX, USA) were used.  We converted the DSL data into IBC values using the conversion formula 
$$
2.02*DSL = IBC 
$$
which has a reported $r^2$ of 0.85 \cite{Hehenkamp2006}, and censored 39 datapoints over 54.3 years (mean age at menopause plus one standard deviation \cite{Treloar81}.)  The resulting dataset consists of  3,260 serum AMH concentrations at known ages, and approximates circulating AMH concentrations in the healthy population from conception to menopause.

\subsection*{Data analysis}

Stratified 10-fold cross-validation was performed using standard techniques \cite{HTF2001}.  The dataset was split into 10 distinct subsets, $k_0, \ldots, k_9$, of nearly equal size, each with similar mean, median, minimum and maximum. For fold $i$, $k_i$ was retained as test data, with the remaining 90\% of the data used for training purposes. 
For each training set we added zero AMH values at conception, in order to force models through the only known AMH concentration at any age. Since variability increases with AMH concentration, we log-adjusted the data (after adding one to each value so that zero AMH on a chart represents zero serum AMH).   All data  were analysed as cross-sectional values (i.e. longitudinal patterns were not considered).  We then fitted 215 mathematical models to the $i$-th test data using TableCurve-2D (Systat Software Inc., San Jose, California, USA), and ranked the results by coefficient of determination, $r^2$. Each model defines a generic type of curve and has parameters which, when instantiated gives a specific curve of that type. For each model we calculated values for the parameters that maximise the $r^2$ coefficient. The Levenberg-Marquardt non-linear curve-fitting algorithm was used throughout, with convergence to 5 significant figures after a maximum of 1,000 iterations.   The highest ranked model was chosen as the best model for the test data, and the mean square error and $r^2$ were calculated after removing the artificial zero values at conception. 

\begin{table}[!ht]
\caption{
\bf{10-fold cross-validation models}}
{\footnotesize
\begin{tabular}{|c|r|r|r|r|r|r|r|r|r|r|r|}
\hline
  & $k_0$ & $k_1$ &$k_2$ &$k_3$& $k_4$ & $k_5$ & $k_6$ &$k_7$ & $k_8$ & $k_9$ & All data\\ 
\hline

$c_0$	&	2.70e-01	&	2.67e-01	&	2.67e-01	&	2.67e-01	&	2.72e-01	&	2.69e-01	&	2.67e-01	&	2.71e-01	&	2.70e-01	&	2.68e-01	&	2.69e-01	\\
$c_1$	&	2.78e-01	&	2.91e-01	&	2.91e-01	&	2.75e-01	&	2.84e-01	&	2.83e-01	&	2.62e-01	&	2.50e-01	&	2.79e-01	&	2.69e-01	&	2.77e-01	\\
$c_2$	&	-1.92e-01	&	-1.90e-01	&	-1.90e-01	&	-1.91e-01	&	-1.98e-01	&	-1.78e-01	&	-1.87e-01	&	-1.91e-01	&	-1.86e-01	&	-1.74e-01	&	-1.88e-01	\\
$c_3$	&	-3.21e-02	&	-4.82e-02	&	-4.82e-02	&	-3.27e-02	&	-3.59e-02	&	-3.42e-02	&	-1.54e-02	&	-2.93e-04	&	-2.95e-02	&	-1.66e-02	&	-2.99e-02	\\
$c_4$	&	7.17e-02	&	8.45e-02	&	8.45e-02	&	7.22e-02	&	7.92e-02	&	6.39e-02	&	5.99e-02	&	4.72e-02	&	6.64e-02	&	5.14e-02	&	6.85e-02	\\
$c_5$	&	-3.13e-02	&	-3.62e-02	&	-3.62e-02	&	-3.14e-02	&	-3.52e-02	&	-2.64e-02	&	-2.78e-02	&	-2.23e-02	&	-2.86e-02	&	-2.27e-02	&	-3.00e-02	\\
$c_6$	&	7.56e-03	&	8.71e-03	&	8.71e-03	&	7.57e-03	&	8.66e-03	&	6.14e-03	&	6.98e-03	&	5.53e-03	&	6.85e-03	&	5.55e-03	&	7.25e-03	\\
$c_7$	&	-1.21e-03	&	-1.38e-03	&	-1.38e-03	&	-1.20e-03	&	-1.40e-03	&	-9.56e-04	&	-1.15e-03	&	-8.98e-04	&	-1.09e-03	&	-9.02e-04	&	-1.16e-03	\\
$c_8$	&	1.38e-04	&	1.57e-04	&	1.57e-04	&	1.37e-04	&	1.61e-04	&	1.07e-04	&	1.35e-04	&	1.03e-04	&	1.23e-04	&	1.05e-04	&	1.33e-04	\\
$c_9$	&	-1.17e-05	&	-1.31e-05	&	-1.31e-05	&	-1.15e-05	&	-1.36e-05	&	-8.90e-06	&	-1.16e-05	&	-8.83e-06	&	-1.04e-05	&	-9.08e-06	&	-1.12e-05	\\
$c_{10}$	&	7.50e-07	&	8.35e-07	&	8.35e-07	&	7.33e-07	&	8.75e-07	&	5.66e-07	&	7.61e-07	&	5.72e-07	&	6.64e-07	&	5.97e-07	&	7.20e-07	\\
$c_{11}$	&	-3.71e-08	&	-4.08e-08	&	-4.08e-08	&	-3.60e-08	&	-4.31e-08	&	-2.78e-08	&	-3.83e-08	&	-2.85e-08	&	-3.28e-08	&	-3.02e-08	&	-3.56e-08	\\
$c_{12}$	&	1.43e-09	&	1.55e-09	&	1.55e-09	&	1.38e-09	&	1.65e-09	&	1.06e-09	&	1.49e-09	&	1.10e-09	&	1.26e-09	&	1.19e-09	&	1.37e-09	\\
$c_{13}$	&	-4.27e-11	&	-4.57e-11	&	-4.57e-11	&	-4.09e-11	&	-4.89e-11	&	-3.16e-11	&	-4.50e-11	&	-3.30e-11	&	-3.78e-11	&	-3.63e-11	&	-4.08e-11	\\
$c_{14}$	&	9.89e-13	&	1.05e-12	&	1.05e-12	&	9.41e-13	&	1.12e-12	&	7.31e-13	&	1.05e-12	&	7.68e-13	&	8.76e-13	&	8.56e-13	&	9.44e-13	\\
$c_{15}$	&	-1.76e-14	&	-1.83e-14	&	-1.83e-14	&	-1.66e-14	&	-1.98e-14	&	-1.29e-14	&	-1.88e-14	&	-1.37e-14	&	-1.56e-14	&	-1.55e-14	&	-1.67e-14	\\
$c_{16}$	&	2.34e-16	&	2.41e-16	&	2.41e-16	&	2.21e-16	&	2.61e-16	&	1.72e-16	&	2.52e-16	&	1.83e-16	&	2.09e-16	&	2.10e-16	&	2.23e-16	\\
$c_{17}$	&	-2.27e-18	&	-2.31e-18	&	-2.31e-18	&	-2.13e-18	&	-2.51e-18	&	-1.67e-18	&	-2.46e-18	&	-1.78e-18	&	-2.03e-18	&	-2.06e-18	&	-2.16e-18	\\
$c_{18}$	&	1.51e-20	&	1.52e-20	&	1.52e-20	&	1.41e-20	&	1.65e-20	&	1.11e-20	&	1.64e-20	&	1.19e-20	&	1.36e-20	&	1.39e-20	&	1.43e-20	\\
$c_{19}$	&	-6.16e-23	&	-6.12e-23	&	-6.12e-23	&	-5.74e-23	&	-6.66e-23	&	-4.54e-23	&	-6.74e-23	&	-4.84e-23	&	-5.55e-23	&	-5.75e-23	&	-5.83e-23	\\
$c_{20}$	&	1.16e-25	&	1.14e-25	&	1.14e-25	&	1.08e-25	&	1.24e-25	&	8.56e-26	&	1.27e-25	&	9.14e-26	&	1.05e-25	&	1.10e-25	&	1.10e-25	\\

   \hline
\end{tabular}
}
\begin{flushleft}  The columns are the 21 coefficients $c_j$ for the degree 20 polynomial returned by TableCurve2D that gave the highest $r^2$ for the dataset under consideration. For  fold $k_i$, the dataset consisted of all points except the $k_i$ test data. After validation of the  degree 20 polynomial  (Table 3), its coefficients were calculated for the entire dataset (final column).  We report this model as our validated model of serum AMH from conception to menopause. 
\end{flushleft}
\label{tab:label}
 \end{table}

For validation purposes, the mean square error of the $k_i$ data for the $i$-th model was calculated and compared to the mean square error of training data for the  same model. In other words, the cross-validation estimate of the prediction  error of the model was compared to the training error of the model. Training error is expected to be underestimated (due to overfitting of the chosen data) and the  cross-validation estimate of the prediction  error  is expected to be overestimated (due to underfitting of unseen data). We consider a model to be validated if the differences between these errors is small. 

For each fold the highest ranked model was a degree 20 polynomial of the form
$$
log_{10}(AMH + 1) = c_0 + c_1\mathrm{age}  + c_2\mathrm{age}^2 + \ldots +  c_{20}\mathrm{age}^{20}
$$
Hence, after validation, the  21 parameters for this model were derived for the entire dataset (n = 3,260), again using TableCurve2D. The resulting model is reported as our validated model (Table 2). 

\begin{table}[!ht]
\caption{
\bf{10-fold cross-validation results}}
\begin{tabular}{|c||r|r|r|r|r|r|r|r|r|r||r|}
\hline
  & $k_0$ & $k_1$ &$k_2$ &$k_3$& $k_4$ & $k_5$ & $k_6$ &$k_7$ & $k_8$ & $k_9$ & Mean\\ 
\hline
 MSE train & 0.069 &	0.068&	0.067&	0.069&	0.069&	0.069&	0.069&	0.069&	0.069&	0.068&		\bf{0.069}\\
 MSE test & 0.064&	0.070&	0.080&	0.065&	0.068&	0.069&	0.067&	0.068&	0.064&	0.078&		\bf{0.069}\\
 $r^2$ & 0.338&	0.338&	0.341&	0.352&	0.343&	0.339&	0.341&	0.340&	0.337&	0.345&		\bf{0.341}\\
 Age at peak AMH & 25.1	&21.7&	21.5&	22.0&	25.5&	24.5&	24.6&	24.5	&24.1&	25.0	&	\bf{23.8}\\
                  \hline
\end{tabular}
\begin{flushleft} The mean squared error (MSE) for the model with highest $r^2$ is given for both the training set (90\% of the dataset) and the test set (the remaining 10\%) for each of the ten folds. Since -- both for individual folds and on average -- the errors are similar,  we consider the model to be validated. The $r^2$ and peak ages are for the highest ranked model returned by TableCurve2D for each fold. 
\end{flushleft}
\label{tab:label}
 \end{table}


\section*{Acknowledgments}

\bibliography{AMHData}

\begin{thebibliography}{10}
\providecommand{\url}[1]{\texttt{#1}}
\providecommand{\urlprefix}{URL }
\expandafter\ifx\csname urlstyle\endcsname\relax
  \providecommand{\doi}[1]{doi:\discretionary{}{}{}#1}\else
  \providecommand{\doi}{doi:\discretionary{}{}{}\begingroup
  \urlstyle{rm}\Url}\fi
\providecommand{\bibAnnoteFile}[1]{%
  \IfFileExists{#1}{\begin{quotation}\noindent\textsc{Key:} #1\\
  \textsc{Annotation:}\ \input{#1}\end{quotation}}{}}
\providecommand{\bibAnnote}[2]{%
  \begin{quotation}\noindent\textsc{Key:} #1\\
  \textsc{Annotation:}\ #2\end{quotation}}
\providecommand{\eprint}[2][]{\url{#2}}

\bibitem{Broekmans09}
Broekmans FJ, Soules MR, Fauser BC (2009) {{O}varian aging: mechanisms and
  clinical consequences}.
\newblock Endocrine Reviews 30: 465--493.
\bibAnnoteFile{Broekmans09}

\bibitem{Gougeon96}
Gougeon A (1996) {{R}egulation of ovarian follicular development in primates:
  facts and hypotheses}.
\newblock Endocrine Reviews 17: 121--155.
\bibAnnoteFile{Gougeon96}

\bibitem{McGee00}
McGee EA, Hsueh AJ (2000) {{I}nitial and cyclic recruitment of ovarian
  follicles}.
\newblock Endocrine Reviews 21: 200--214.
\bibAnnoteFile{McGee00}

\bibitem{Durlinger02}
Durlinger AL, Visser JA, Themmen AP (2002) {{R}egulation of ovarian function:
  the role of anti-{M}\"{u}llerian hormone}.
\newblock Reproduction 124: 601--609.
\bibAnnoteFile{Durlinger02}

\bibitem{Gigli05}
Gigli I, Cushman RA, Wahl CM, Fortune JE (2005) {{E}vidence for a role for
  anti-{M}ullerian hormone in the suppression of follicle activation in mouse
  ovaries and bovine ovarian cortex grafted beneath the chick chorioallantoic
  membrane}.
\newblock Molecular Reproductive Development 71: 480--488.
\bibAnnoteFile{Gigli05}

\bibitem{Durlinger99}
Durlinger AL, Kramer P, Karels B, de~Jong FH, Uilenbroek JT, et~al. (1999)
  {{C}ontrol of primordial follicle recruitment by anti-{M}\"{u}llerian hormone
  in the mouse ovary}.
\newblock Endocrinology 140: 5789--5796.
\bibAnnoteFile{Durlinger99}

\bibitem{Durlinger01}
Durlinger AL, Gruijters MJ, Kramer P, Karels B, Kumar TR, et~al. (2001)
  {{A}nti-{M}\"{u}llerian hormone attenuates the effects of {F}{S}{H} on
  follicle development in the mouse ovary}.
\newblock Endocrinology 142: 4891--4899.
\bibAnnoteFile{Durlinger01}

\bibitem{Weenen04}
Weenen C, Laven JS, Von~Bergh AR, Cranfield M, Groome NP, et~al. (2004)
  {{A}nti-{M}\"{u}llerian hormone expression pattern in the human ovary:
  potential implications for initial and cyclic follicle recruitment}.
\newblock Molecular Human Reproduction 10: 77--83.
\bibAnnoteFile{Weenen04}

\bibitem{Bezard87}
Bezard J, Vigier B, Tran D, Mauleon P, Josso N (1987) {{I}mmunocytochemical
  study of anti-{M}\"{u}llerian hormone in sheep ovarian follicles during fetal
  and post-natal development}.
\newblock Journal of Reproduction and Fertility 80: 509--516.
\bibAnnoteFile{Bezard87}

\bibitem{Baarends95}
Baarends WM, Uilenbroek JT, Kramer P, Hoogerbrugge JW, van Leeuwen EC, et~al.
  (1995) {{A}nti-{m}\"{u}llerian hormone and anti-m\"{u}llerian hormone type
  {I}{I} receptor messenger ribonucleic acid expression in rat ovaries during
  postnatal development, the estrous cycle, and gonadotropin-induced follicle
  growth}.
\newblock Endocrinology 136: 4951--4962.
\bibAnnoteFile{Baarends95}

\bibitem{Hansen10}
Hansen KR, Hodnett GM, Knowlton N, Craig LB (2011) {{C}orrelation of ovarian
  reserve tests with histologically determined primordial follicle number}.
\newblock Fertility \& Sterility 95: 170--175.
\bibAnnoteFile{Hansen10}

\bibitem{Kevenaar06}
Kevenaar ME, Meerasahib MF, Kramer P, van~de Lang-Born BM, de~Jong FH, et~al.
  (2006) {{S}erum anti-{m}\"{u}llerian hormone levels reflect the size of the
  primordial follicle pool in mice}.
\newblock Endocrinology 147: 3228--3234.
\bibAnnoteFile{Kevenaar06}

\bibitem{Lintern-Moore74}
Lintern-Moore S, Peters H, Moore GP, Faber M (1974) {{F}ollicular development
  in the infant human ovary}.
\newblock Journal of Reproduction and Fertility 39: 53--64.
\bibAnnoteFile{Lintern-Moore74}

\bibitem{WK}
Wallace WHB, Kelsey TW (2010) {{H}uman ovarian reserve from conception to the
  menopause}.
\newblock PLoS ONE 5: e8772.
\bibAnnoteFile{WK}

\bibitem{Faddy92}
Faddy MJ, Gosden RG, Gougeon A, Richardson SJ, Nelson JF (1992) {{A}ccelerated
  disappearance of ovarian follicles in mid-life: implications for forecasting
  menopause}.
\newblock Human Reproduction 7: 1342--1346.
\bibAnnoteFile{Faddy92}

\bibitem{Faddy96}
Faddy MJ, Gosden RG (1996) {{A} model conforming the decline in follicle
  numbers to the age of menopause in women}.
\newblock Human Reproduction 11: 1484--1486.
\bibAnnoteFile{Faddy96}

\bibitem{Faddy00}
Faddy MJ (2000) {{F}ollicle dynamics during ovarian ageing}.
\newblock Molecular \& Cellular Endocrinology 163: 43--48.
\bibAnnoteFile{Faddy00}

\bibitem{Hansen08}
Hansen KR, Knowlton NS, Thyer AC, Charleston JS, Soules MR, et~al. (2008) A new
  model of reproductive aging: the decline in ovarian non-growing follicle
  number from birth to menopause.
\newblock Human Reproduction 23: 699-708.
\bibAnnoteFile{Hansen08}

\bibitem{DeVet2002}
de~Vet A, Laven J, {De Jong} F, Themmen APN, Fauser BCJM (2002)
  {Antim\"{u}llerian hormone serum levels: a putative marker for ovarian
  aging}.
\newblock Fertility \& Sterility 77: 357--362.
\bibAnnoteFile{DeVet2002}

\bibitem{VanRooij2005}
van Rooij IAJ, Broekmans FJM, Scheffer GJ, Looman CWN, Habbema JDF, et~al.
  (2005) {Serum antim\"{u}llerian hormone levels best reflect the reproductive
  decline with age in normal women with proven fertility: a longitudinal
  study.}
\newblock Fertility \& Sterility 83: 979--87.
\bibAnnoteFile{VanRooij2005}

\bibitem{Cook00}
Cook CL, Siow Y, Taylor S, Fallat ME (2000) {{S}erum m\"{u}llerian-inhibiting
  substance levels during normal menstrual cycles}.
\newblock Fertility \& Sterility 73: 859--861.
\bibAnnoteFile{Cook00}

\bibitem{LaMarca04}
La~Marca A, Malmusi S, Giulini S, Tamaro LF, Orvieto R, et~al. (2004)
  {{A}nti-{M}\"{u}llerian hormone plasma levels in spontaneous menstrual cycle
  and during treatment with {F}{S}{H} to induce ovulation}.
\newblock Human Reproduction 19: 2738--2741.
\bibAnnoteFile{LaMarca04}

\bibitem{Fanchin05}
Fanchin R, Taieb J, Lozano DH, Ducot B, Frydman R, et~al. (2005) {{H}igh
  reproducibility of serum anti-{M}ullerian hormone measurements suggests a
  multi-staged follicular secretion and strengthens its role in the assessment
  of ovarian follicular status}.
\newblock Human Reproduction 20: 923--927.
\bibAnnoteFile{Fanchin05}

\bibitem{LaMarca10}
La~Marca A, Sighinolfi G, Radi D, Argento C, Baraldi E, et~al. (2010)
  {{A}nti-{M}ullerian hormone ({A}{M}{H}) as a predictive marker in assisted
  reproductive technology ({A}{R}{T})}.
\newblock Human Reproduction Update 16: 113--130.
\bibAnnoteFile{LaMarca10}

\bibitem{Nelson10}
Nelson SM, Messow MC, Wallace AM, Fleming R, McConnachie A (2011) {{N}omogram
  for the decline in serum antim\"{u}llerian hormone: a population study of
  9,601 infertility patients}.
\newblock Fertility \& Sterility 95: 736--741.
\bibAnnoteFile{Nelson10}

\bibitem{Nelson11}
Nelson SM, Messow MC, McConnachie A, Wallace WHB, Kelsey TW, et~al. (2011)
  {External validation of nomogram for the decline in serum anti-M\"{u}llerian
  hormone in women: a population study of 15,834 infertility patients}.
\newblock Reproductive BioMedicine Online (published online 20 May 2011).
\bibAnnoteFile{Nelson11}

\bibitem{Seifer11}
Seifer DB, Baker VL, Leader B (2011) {{A}ge-specific serum anti-{M}\"{u}llerian
  hormone values for 17,120 women presenting to fertility centers within the
  {U}nited {S}tates}.
\newblock Fertility \& Sterility 95: 747--750.
\bibAnnoteFile{Seifer11}

\bibitem{Almog11}
Almog B, Shehata F, Suissa S, Holzer H, Shalom-Paz E, et~al. (2011)
  {{A}ge-related normograms of serum antim\"{u}llerian hormone levels in a
  population of infertile women: a multicenter study}.
\newblock Fertility \& Sterility (published online 20 May 2011).
\bibAnnoteFile{Almog11}

\bibitem{Hagen2010}
Hagen CP, Aksglaede L, Sorensen K, Main KM, Boas M, et~al. (2010) {{S}erum
  levels of anti-{M}\"{u}llerian hormone as a marker of ovarian function in 926
  healthy females from birth to adulthood and in 172 {T}urner syndrome
  patients}.
\newblock Journal of Clinical Endocrinology \& Metababolism 95: 5003--5010.
\bibAnnoteFile{Hagen2010}

\bibitem{Ahmed2010}
Ahmed SF, Keir L, McNeilly J, Galloway P, O'Toole S, et~al. (2010) {The
  concordance between serum anti-M\"{u}llerian hormone and testosterone
  concentrations depends on duration of hCG stimulation in boys undergoing
  investigation of gonadal function.}
\newblock Clinical Endocrinology 72: 814--9.
\bibAnnoteFile{Ahmed2010}

\bibitem{Lee03}
Lee MM (2003) {{R}eproductive hormones in infant girls--a harbinger of adult
  reproductive function?}
\newblock Journal of Clinical Endocrinology \& Metabolism 88: 3513--3514.
\bibAnnoteFile{Lee03}

\bibitem{Chellakooty03}
Chellakooty M, Schmidt IM, Haavisto AM, Boisen KA, Damgaard IN, et~al. (2003)
  {{I}nhibin {A}, inhibin {B}, follicle-stimulating hormone, luteinizing
  hormone, estradiol, and sex hormone-binding globulin levels in 473 healthy
  infant girls}.
\newblock Journal of Clinical Endocrinology \& Metabolism 88: 3515--3520.
\bibAnnoteFile{Chellakooty03}

\bibitem{Andersson98}
Andersson AM, Toppari J, Haavisto AM, Petersen JH, Simell T, et~al. (1998)
  {{L}ongitudinal reproductive hormone profiles in infants: peak of inhibin {B}
  levels in infant boys exceeds levels in adult men}.
\newblock Journal of Clinical Endocrinology \& Metabolism 83: 675--681.
\bibAnnoteFile{Andersson98}

\bibitem{VanDisseldorp2008}
van Disseldorp J, Faddy MJ, Themmen aPN, de~Jong FH, Peeters PHM, et~al. (2008)
  {Relationship of serum antim\"{u}llerian hormone concentration to age at
  menopause.}
\newblock Journal of Clinical Endocrinology \& Metabolism 93: 2129--34.
\bibAnnoteFile{VanDisseldorp2008}

\bibitem{Soto2009}
Soto N, I\~{n}iguez G, L\'{o}pez P, Larenas G, Mujica V, et~al. (2009)
  {Anti-M\"{u}llerian hormone and inhibin B levels as markers of premature
  ovarian aging and transition to menopause in type 1 diabetes mellitus.}
\newblock Human Reproduction 24: 2838--44.
\bibAnnoteFile{Soto2009}

\bibitem{LaMarca2005}
{La Marca} A, {De Leo} V, Giulini S, Orvieto R, Malmusi S, et~al. (2005)
  {Anti-M\"{u}llerian hormone in premenopausal women and after spontaneous or
  surgically induced menopause.}
\newblock Journal of the Society for Gynecologic Investigation 12: 545--8.
\bibAnnoteFile{LaMarca2005}

\bibitem{Sowers09}
Sowers M, McConnell D, Gast K, Zheng H, Nan B, et~al. (2010)
  {{A}nti-{M}\"{u}llerian hormone and inhibin {B} variability during normal
  menstrual cycles}.
\newblock Fertility \& Sterility 94: 1482--1486.
\bibAnnoteFile{Sowers09}

\bibitem{Guibourdenche2003}
Guibourdenche J, Lucidarme N, Chevenne D, Rigal O, Nicolas M, et~al. (2003)
  {Anti-M\"{u}llerian hormone levels in serum from human foetuses and children:
  pattern and clinical interest}.
\newblock Molecular and Cellular Endocrinology 211: 55--63.
\bibAnnoteFile{Guibourdenche2003}

\bibitem{Hudecova2009}
Hudecova M, Holte J, Olovsson M, {Sundstr\"{o}m Poromaa} I (2009) {Long-term
  follow-up of patients with polycystic ovary syndrome: reproductive outcome
  and ovarian reserve.}
\newblock Human Reproduction 24: 1176--83.
\bibAnnoteFile{Hudecova2009}

\bibitem{Mulders2004}
Mulders AGMGJ, Laven JSE, Eijkemans MJC, de~Jong FH, Themmen APN, et~al. (2004)
  {Changes in anti-M\"{u}llerian hormone serum concentrations over time suggest
  delayed ovarian ageing in normogonadotrophic anovulatory infertility.}
\newblock Human Reproduction 19: 2036--42.
\bibAnnoteFile{Mulders2004}

\bibitem{Pastor2005}
Pastor CL, Vanderhoof VH, Lim LCL, Calis KA, Premkumar A, et~al. (2005) {Pilot
  study investigating the age-related decline in ovarian function of regularly
  menstruating normal women.}
\newblock Fertility \& Sterility 84: 1462--9.
\bibAnnoteFile{Pastor2005}

\bibitem{Piltonen2005}
Piltonen T, Morin-Papunen L, Koivunen R, Perheentupa A, Ruokonen A, et~al.
  (2005) {Serum anti-M\"{u}llerian hormone levels remain high until late
  reproductive age and decrease during metformin therapy in women with
  polycystic ovary syndrome.}
\newblock Human Reproduction 20: 1820--6.
\bibAnnoteFile{Piltonen2005}

\bibitem{Laven2004}
Laven JSE (2004) {Anti-M\"{u}llerian Hormone Serum Concentrations in
  Normoovulatory and Anovulatory Women of Reproductive Age}.
\newblock Journal of Clinical Endocrinology \& Metabolism 89: 318--323.
\bibAnnoteFile{Laven2004}

\bibitem{Knauff2009}
Knauff EaH, Eijkemans MJC, Lambalk CB, ten Kate-Booij MJ, Hoek A, et~al. (2009)
  {Anti-M\"{u}llerian hormone, inhibin B, and antral follicle count in young
  women with ovarian failure.}
\newblock Journal of Clinical Endocrinology \& Metabolism 94: 786--92.
\bibAnnoteFile{Knauff2009}

\bibitem{Lee1996}
Lee MM, Donahoe PK, Hasegawa T, Silverman B, Crist GB, et~al. (1996)
  {M\"{u}llerian inhibiting substance in humans: normal levels from infancy to
  adulthood.}
\newblock Journal of Clinical Endocrinology \& Metabolism 81: 571--6.
\bibAnnoteFile{Lee1996}

\bibitem{VanBeek2007}
van Beek RD, van~den Heuvel-Eibrink MM, Laven JSE, de~Jong FH, Themmen APN,
  et~al. (2007) {Anti-M\"{u}llerian hormone is a sensitive serum marker for
  gonadal function in women treated for Hodgkin's lymphoma during childhood.}
\newblock Journal of Clinical Endocrinology \& Metabolism 92: 3869--74.
\bibAnnoteFile{VanBeek2007}

\bibitem{Sanders2009}
Sanders RD, Spencer JB, Epstein MP, Pollak SV, Vardhana PA, et~al. (2009)
  {Biomarkers of ovarian function in girls and women with classic
  galactosemia.}
\newblock Fertility \& Sterility 92: 344--51.
\bibAnnoteFile{Sanders2009}

\bibitem{Tehrani2010}
Tehrani FR, Solaymani-Dodaran M, Hedayati M, Azizi F (2010) {{I}s polycystic
  ovary syndrome an exception for reproductive aging?}
\newblock Human Reproduction 25: 1775--1781.
\bibAnnoteFile{Tehrani2010}

\bibitem{Dorgan2009}
Dorgan JF, Stanczyk FZ, Egleston BL, Kahle LL, Shaw CM, et~al. (2009)
  {Prospective case-control study of serum m\"{u}llerian inhibiting substance
  and breast cancer risk.}
\newblock Journal of the National Cancer Institute 101: 1501--9.
\bibAnnoteFile{Dorgan2009}

\bibitem{PlotDigitizer}
Huwaldt J (2005).
\newblock Plot {D}igitizer.
\newblock \texttt{http://plotdigitizer.sourceforge.net/}.
\bibAnnoteFile{PlotDigitizer}

\bibitem{Hehenkamp2006}
Hehenkamp WJK, Looman CWN, Themmen APN, de~Jong FH, {Te Velde} ER, et~al.
  (2006) {Anti-M\"{u}llerian hormone levels in the spontaneous menstrual cycle
  do not show substantial fluctuation.}
\newblock Journal of Clinical Endocrinology \& Metabolism 91: 4057--63.
\bibAnnoteFile{Hehenkamp2006}

\bibitem{Treloar81}
Treloar AE (1981) {{M}enstrual cyclicity and the pre-menopause}.
\newblock Maturitas 3: 249--264.
\bibAnnoteFile{Treloar81}

\bibitem{HTF2001}
Hastie T, Tibshirani R, Friedman JH (2001) The elements of statistical
  learning: data mining, inference, and prediction.
\newblock New York: Springer-Verlag, 533 pp.
\bibAnnoteFile{HTF2001}

\end{thebibliography}




\end{document}